\documentclass[a4, aps, prb, reprint]{revtex4}

\usepackage{graphicx}
\usepackage{bm}
\usepackage{amsmath}

\begin{document}
Copyright (2013) American Institute of Physics. This article may be downloaded for personal use only. Any other use requires prior permission of the author and the American Institute of Physics. The following article appeared in J. Appl. Phys. 114, 023910 (2013) and may be found at  \url{http://dx.doi.org/10.1063/1.4813228}

\title{Effect of Hole Shape on Spin-Wave Band Structure in One-Dimensional Magnonic Antidot Waveguide}

\author{D. Kumar$^1$, P. Sabareesan$^1$, W. Wang$^2$, H. Fangohr$^2$,}
\author{A. Barman$^1$}
\email{abarman@bose.res.in}
\affiliation{$^1$Thematic Unit of Excellence on Nanodevice Technology and  Department of Condensed Matter Physics and Material Sciences, \\S. N. Bose National Centre for Basic Sciences, Block JD, Sector III, Salt Lake, Kolkata 700 098, India;\\$^2$Engineering and the Environment, University of Southampton, Southampton, SO17 1BJ, UK.}

\date{\today}

\begin{abstract}
We present the possibility of tuning the spin-wave band structure, particularly the bandgaps in a nanoscale magnonic antidot waveguide by varying the shape of the antidots. The effects of changing the shape of the antidots on the spin-wave dispersion relation in a waveguide have been carefully monitored. We interpret the observed variations by analysing the equilibrium magnetic configuration and the magnonic power and phase distribution profiles during spin-wave dynamics. The inhomogeneity in the exchange fields at the antidot boundaries within the waveguide is found to play a crucial role in controlling the band structure at the discussed length scales. The observations recorded here will be important for future developments of magnetic antidot based magnonic crystals and waveguides.

\end{abstract}

\maketitle

\section{Introduction}\label{Sec:I}
Magnonics \cite{Kruglyak2010, Serga2010, Lenk2011, Demokritov2013} is an emerging sub-field of solid state physics, which studies the propagation of spin-waves (SWs) in micro- and nanoscale magnetic structures. Magnonic devices \cite{Khitun2008} aim to use the information carried by SWs to perform their respective designated tasks. Waveguides \cite{Venkat2013}, SW interferometers \cite{Choi2006, Podbielski2006, Schneider2008}, phase shifters \cite{Au2012} and magnonic crystals (MCs) \cite{Lenk2011, Nikitov2001, Neusser2009, Chumak2010} are some of the important components of magnonic devices. Knowledge of spin-wave dispersion within such structures is necessary for their design and operation. An MC can be realized by a combination of periodic modulation of structural and material parameters of a known magnetic material and a control over the external bias magnetic field.\cite{Sykes1976, Bayer2006, Kruglyak2005} These periodic modulations of magnetic potentials within an MC interact with the spin-waves eventually yielding a characteristic dispersion relation comprising of stop and pass bands. Most MCs that form the topic of current research in magnonics are either one-dimensional (1D) \cite{Chumak2008, Wang2009} or two-dimensional (2D) \cite{Mamica2012, Vysotskii2010, Tacchi2010, Tacchi2011, Saha2013, Wang2013} as they are easier to fabricate on a wafer when compared to three-dimensional (3D) MCs. Nevertheless, few theoretical reports on the study of dispersion of SWs in 3D MCs have been made.\cite{Dyson1956, Krawczyk2008}

Structured magnonic waveguides \cite{Lee2009, Kim2009, Ciubotaru2012} have recently attracted considerable attention due to their selective transmission of microwave bands in the micro- and nano-scales and their potential applications in on-chip microwave signal processing and communication. Magnetic antidots have emerged as an important system of MCs; and a thorough investigation of high frequency magnetization dynamics in them have been reported in the literature. \cite{Pechan2005, Neusser2010, Neusser2011, Tacchi2012, Hu2011, Mandal2012} Magnonic antidot waveguide (MAW) is an attractive option for manipulation of transmitted spin waves towards the above application, but it has only recently been started to be explored. \cite{Ma2011, Kumar2012, Klos2012} So far a study of the dependence of spin-wave dispersion on the shapes of the antidots has not been reported. More importantly, how changes in the exchange field distribution around the antidot boundary can alter the characteristic dispersion of exchange or dipole-exchange SWs in a MAW, has never been observed before.

This article aims to help fill that gap in research by numerically simulating the magnonic dispersion in 1D MAW lattices with different geometric shapes of the antidots. We also study the spatial magnetization distribution for different frequencies and wavevectors of the observed dispersion modes. We further plot exchange and demagnetization fields to examine how they change with differing antidot shapes. We have used antidots, which are $n$ sided regular convex polygons inscribed within a circumcircle of radius,\begin{eqnarray}r_n=\sqrt{\frac{2fA}{n}\text{cosec}(\frac{2\pi}{n})};\label{rn}\end{eqnarray} such that, the filling fraction $f$, the ratio of area of the hole to the area $A$ of the unit cell, remains a constant. Micromagnetic simulations were performed for $n=3$ (triangular), $4$ (square), $5$ (pentagonal) and $6$ (hexagonal antidots) in Object-Oriented Micromagnetic Framework (OOMMF).\cite{Donahue2002} The case of $n=\infty$ (circular antidots) was simulated using Nmag.\cite{Fischbacher2007} The paper is organized as follows. The geometrical structure of the waveguide and method used for calculating dispersion are described in greater detail in Sec. \ref{Sec:II}. Section \ref{Sec:III} presents the results and analysis linking the ground state field distribution with changes in the observed SW dispersion modes. Section \ref{Sec:IV} contains the concluding remarks.

\section{MAW and The Numerical Method}\label{Sec:II}

\subsection{MAW Structural and Material Parameters}

\begin{figure}[!ht]
\includegraphics[width=8.5 cm]{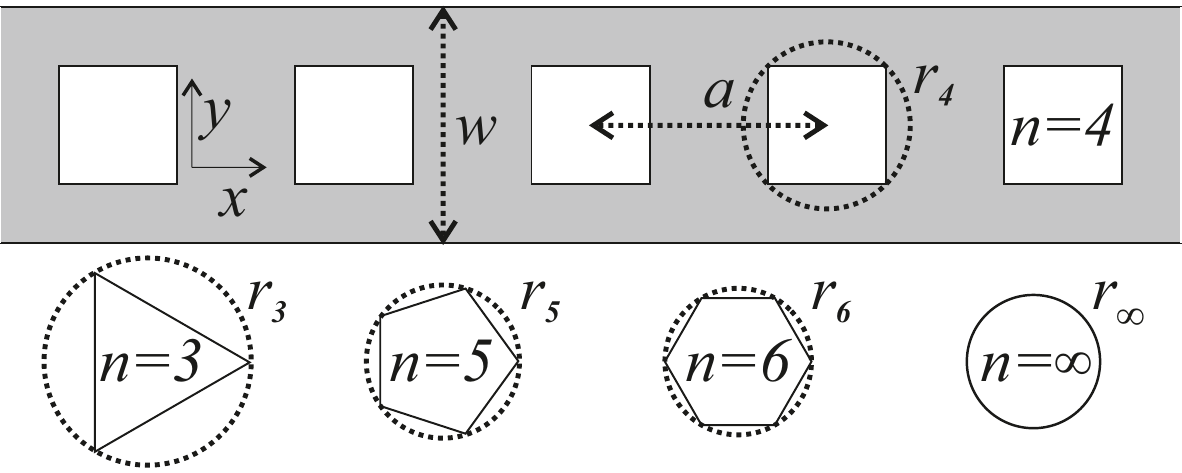}
\caption{(Top panel) A part of the 1D MAW structure showing square antidots (white holes in grey magnetic region) disposed along the central axis of the waveguide of width, $w=24$ nm and lattice constant, $a=24$ nm. The square antidots are inscribed within a circle of radius, $r_4$. (Bottom panel) Other examined antidot shapes inscribed within their respective imaginary circumcircles. For $n\in\{3, 4, 5, 6, \infty\}$, $r_n$ is given by Eqn. \ref{rn}, where filling fraction $f=0.25$ and unit cell area $A=wa$.}
\label{Fig:structure}
\end{figure}

Figure \ref{Fig:structure} depicts the MAW structures under investigations. The MAWs had both width, $w$ and lattice constant, $a$ set to $24$ nm and a length, $l$ and thickness, $s$ of $2.4$ $\mu$m and $3$ nm in all cases. For $f=0.25$, $A=wa$ and $n\in\{3, 4, 5, 6, \infty\}$, Eq. \ref{rn} dictates $r_n$ as $21.06, 16.97, 15.56, 14.89$ and $13.54$ nm, respectively. The material parameters similar to that of permalloy (Py: Ni$_{80}$Fe$_{20}$) were used during simulations (exchange constant, $A=13{\times}10^{-12}$ J/m, saturation magnetization, $M_s=0.8{\times}10^6$ A/m, gyromagnetic ratio, $\bar{\gamma}=2.21{\times}10^5$ m/As and no magnetocrystalline anisotropy).

\subsection{Micromagnetic Simulations}

Micromagnetic simulations \cite{Fidler2000} are done with the help of the finite difference method (FDM) based OOMMF (for $n=3, 4, 5$ and $6$) or the finite element method (FEM) based Nmag (for $n=\infty$). For the cell size used here, Nmag reproduces the circular shape much better than that obtained in OOMMF. The use of two different simulation packages also ensures that the established results are independent of the spatial discretisation. Both these open source platforms solve the Landau-Lifshitz-Gilbert (LLG) equation \cite{Landau1935, Gilbert2004}:

\begin{eqnarray}
\frac{d {\bf M}}{dt}&=&-\bar{\gamma}{\bf M}\times {\bf H}_{\text{eff}}-\frac{\alpha \bar{\gamma}}{M_{\text{s}}}{{\bf M}}\times\left({\bf M}\times {\bf H}_{\text{eff}}\right)\text{.}\label{eq:ll}
\end{eqnarray}

In order to obtain the SW dispersion relations, a 2D discrete Fourier transform (DFT) was performed on the obtained results.\cite{Kumar2012} Before simulating the SW dynamics, a magnetic steady state was achieved by subjecting the MAWs to an external bias of $1.01$ T (along the length of the waveguide) under a Gilbert damping constant, $\alpha=0.95$. This high external field saturates the magnetization of MAWs. To observe sharper dispersion peaks $\alpha$ was artificially reduced to $10^{-4}$ during simulation of the dynamics. For simulations done in OOMMF, cuboidal cells of dimensions $dx=dy=d=1$ nm and $dz=s=3$ nm were used to span the MAWs. The resultant gridding of antidot edges which are not aligned with $X$ or $Y$ axes may cause the entire hole geometry to move towards one of the edges of the MAW. How this intrinsic mirror symmetry breaking affects the SW dispersion relations is the subject of a separate study.\cite{Klos2013} Nmag, being FEM based, uses adaptive meshing and hence, its outputs do not suffer from this issue. However, spatial interpolation needs to be done in order to obtain magnetization values at every $1$ nm interval before performing the DFT. Data was collected every $dt=1$ ps for both OOMMF and Nmag for a total duration of $4$ ns. This gives us a sampling frequency, $f_s=1000$ GHz. The excitation signal, $H_z$ is normal to the plane of the MAWs and is given by:

\begin{eqnarray}
H_z=H_0\left({\frac{\sin(2{\pi}f_{c}(t-t_0)}{2{\pi}f_{c}(t-t_0)}}\right) \times \left(\frac{\sin(2{\pi}k_{c}(x-x_0)}{2{\pi}k_{c}(x-x_0)}\right) \label{eq:signal} 
\times\left(\sum_{i=1}^{w/dy}\sin(i{\pi}y/w)\right). \nonumber
\end{eqnarray}

Here ${\mu}_0H_0=6$ mT, $f_c=490$ GHz, $t_0=1/(f_s-2f_c)=50$ ps, $k_c={\pi}/a$ and $x_0=l/2=1$ $\mu$m. This form of excitation signal will excite both symmetric and antisymmetric modes of the dispersion relations in a width confined MAW. The aliasing associated with DFT is mitigated by the fact that the signal given by Eq. \ref{eq:signal} carries no power beyond $f_c$ in the frequency domain. Similarly, power in the wavevector domain is limited to the first Brillouin zone (BZ) from $-k_c$ to $k_c$.

We also calculated the SW power and phase distribution profiles (PPDPs) for a given ($k$, $f$) pair of any dispersion relation. It was done by masking the obtained relation with a suitable mask in wavevector domain followed by doing an inverse Fourier transform in the same domain to yield data in physical space. For example, in order to obtain these results for ($k$, $f$) = ($K$, $F$) a mask, $D_{\text{m}}$ was created to span the entire $k$ vs. $f$ space such that:

\begin{eqnarray}
D_{\text{m}}(k, f)=
\begin{cases}
1 & \text{ if } k=2c\pi/a{\pm}K\text{: }c\text{ is an integer}\\
0 & \text{ elsewhere.}
\end{cases}
\label{eq:mask}
\end{eqnarray}

After multiplying $D_{\text{m}}$ with the obtained dispersion relations we then take an inverse Fourier transform in $k$-space to arrive at the desired PPDPs. This mask is designed to include power only from $k=K$ and nullify the power present in the rest of the wavevector domain. Simply performing the inverse transform in $k$-space without using such a mask will allow power from the entire wavevector range to distort the results.

\section{Results and Observations}\label{Sec:III}

\begin{figure}[!ht]
\includegraphics[width=8.5 cm]{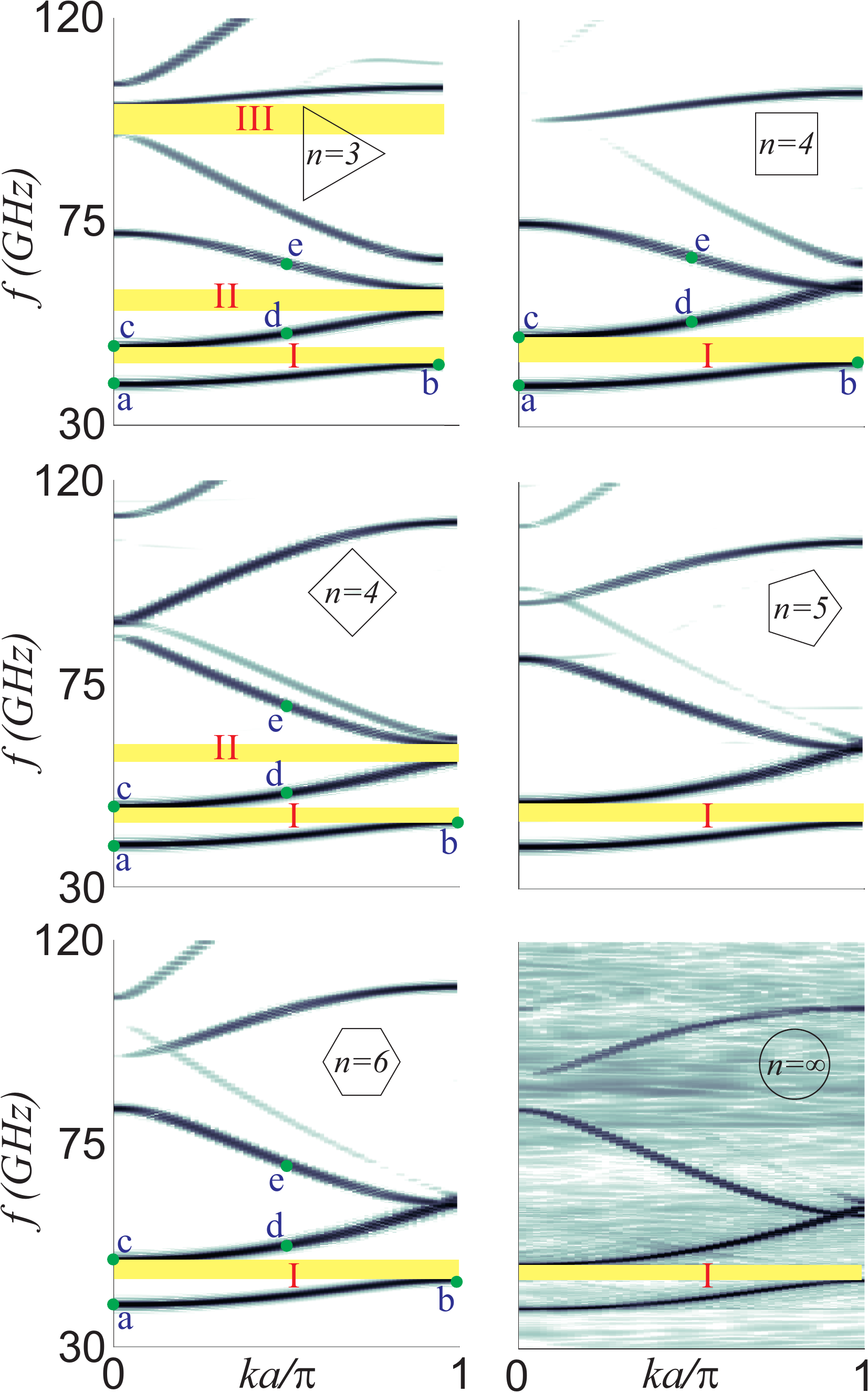}
\caption{(Colour Online) SW dispersion results of MAW structures marked with their respective antidot shapes as insets. Indexed band gaps are highlighted with horizontal bars.}
\label{Fig:dispersion}
\end{figure}

The calculated dispersion relations are tabulated in Fig. \ref{Fig:dispersion}. Frequency ranges from $0$ to $120$ GHz and wavevector $k$ ranges from $0$ to the first BZ boundary ($\pi/a$) are displayed. As the bias field is kept constant at $1.01$ T, a forbidden region is observed in all the cases up to the ferromagnetic resonance mode of about $39$ GHz. SW of any $k$ is not allowed in this region. Bandgap I is also present in all the cases. For triangular, square, pentagonal, hexagonal and circular antidots, its respective values are $4.3$ GHz ($43$ GHz to $47.3$ GHz), $5.6$ GHz ($44.1$ GHz to $49.7$ GHz), $4.4$ GHz ($44.5$ GHz to $48.9$ GHz), $4.4$ GHz ($44.8$ GHz to $49.2$ GHz) and $3.5$ GHz ($44.9$ GHz to $48.4$ GHz). In the case where the square antidots were tilted by $45^{\circ}$ (diamond shaped antidots), bandgaps I \& II were observed; and their respective values were $3.6$ GHz ($44.2$ GHz to $47.8$ GHz) and $3.5$ GHz ($57.8$ GHz to $61.3$ GHz). An additional bandgap (III) of $6.6$ GHz ($94$ GHz to $100.6$ GHz) was observed in the case of triangular antidots. Bandgaps II \& III are direct but bandgap I is indirect suggesting a difference in their origin which can be studied by looking at the spatial PPDPs for the modes between which they exist.

\begin{figure}[!ht]
\includegraphics[width=16 cm]{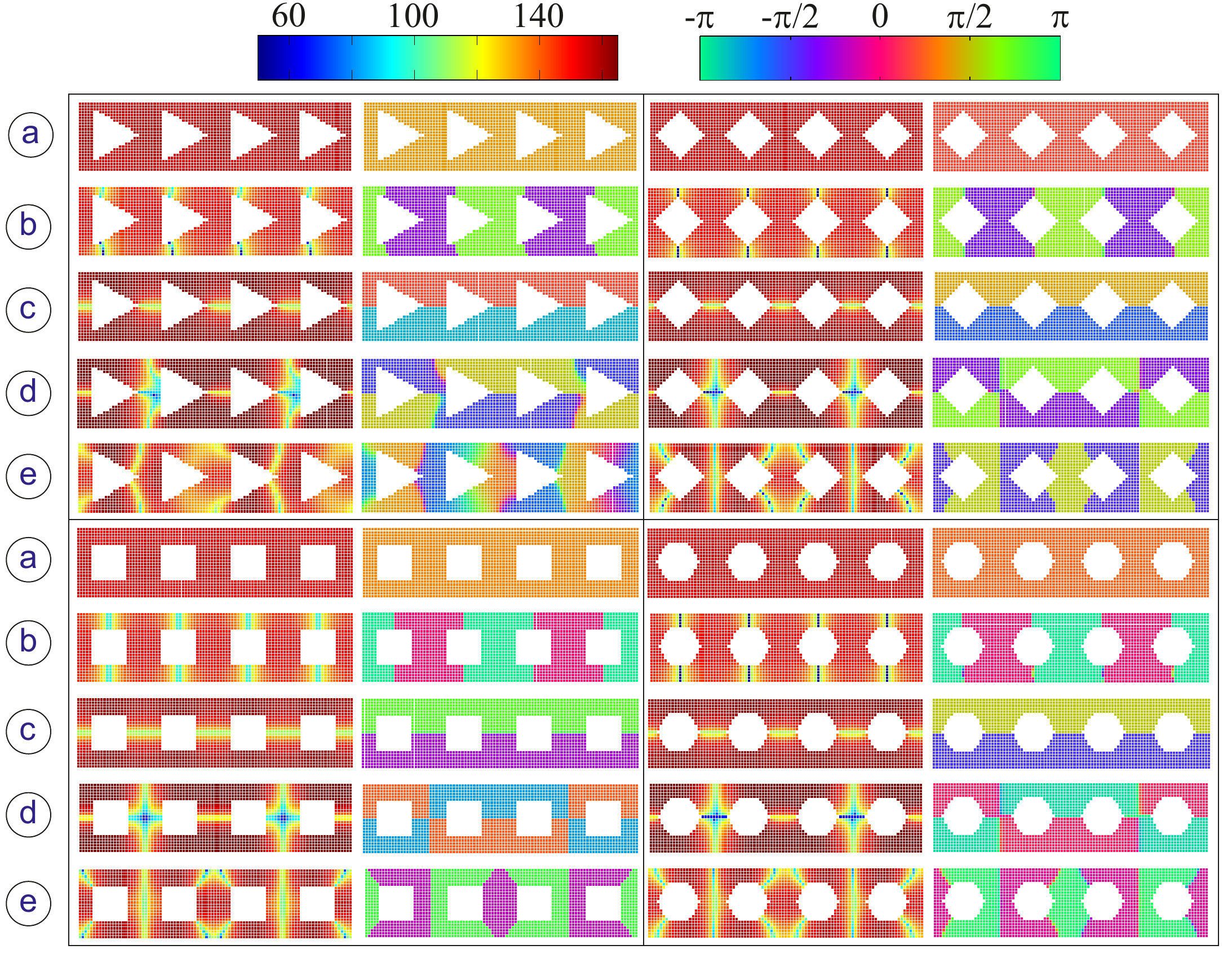}
\caption{(Colour Online) Power (first and third column) and phase (second and fourth column) distribution profiles corresponding to marked $(k,f)$ locations (\textcircled{a} to \textcircled{e}) in Fig. \ref{Fig:dispersion} for MAWs with triangular, diamond, square and hexagonal shaped antidots. Power is presented on an arbitrary logarithmic colour map while the phase profile representations use a cyclic colour map.}
\label{Fig:powerPhase}
\end{figure}

Figure ~\ref{Fig:powerPhase} shows the spatial SW PPDPs for the marked ($k$, $f$) values in the Fig. \ref{Fig:dispersion}. Only a part of the entire MAW structures have been shown for convenience. Mode \textcircled{a} appears to describe the uniform mode showing insignificant power or phase variation in the medium. The power distribution profile (PoDP) of mode \textcircled{b}, being at the BZ boundary, features narrow vertical nodal lines at $x=x_0{\pm}(c+1/2)a$; where $c$ is an integer. The regions joining these nodal lines are $\pi$ radians out of phase with each other. This suggests that the positions of the phase boundaries in the phase distribution profiles (PhDP) depend on the location of the signal $x_0$ used in Eq. \ref{eq:signal}. Power distribution profiles for mode \textcircled{c} contains a horizontal nodal line right down the centre of the MAWs in all cases. The upper and lower parts of the waveguide are again $\pi$ radians out of phase with each other. This hints at the fact that modes \textcircled{a} and \textcircled{c} correspond to zero and first order modes along the width due to the lateral confinement of the waveguide.\cite{Guslienko2002} Modes \textcircled{d} and \textcircled{e} are calculated at $k=\pi/2a$ as they become nearly degenerate at the BZ boundary for square and hexagonal antidots. This degeneracy can lead one of the modes to effect the results of the other. Vertical nodal lines for both these modes are now located at $x=x_0{\pm}(2c+1)a$. Yet again,
the positions of the phase boundaries appear to be controlled by the location of the signal at $x_0$. The periodicity of these nodal lines $2a$ is understandable given the location of modes (half way from BZ boundary). Slight curvature is observed in all the nodal lines for triangular antidots. We attribute this to the lack of mirror symmetry within the hole geometry along a vertical axis. Similar curvature of nodal lines was detected for the MAW with pentagonal antidots (not shown) which also lacked such a symmetry. Belonging to the same dispersive branch of the spectrum, modes \textcircled{c} and \textcircled{d} share a horizontal nodal line which stems from the aforementioned lateral confinement. The observed effects of such confinement and the shape of dispersion curve to which modes \textcircled{a} through \textcircled{d} belong reminds us of the first two (nearly) parabolic dispersion curves observed in the case of a uniform waveguide.\cite{Venkat2013} In contrast, mode \textcircled{e} belongs to dispersive branch in the spectrum, which curves downwards. This branch is formed by the anti-crossing of lowest energy modes originating in the two neighbours of a BZ; and as such mode \textcircled{e} unlike modes \textcircled{a} and \textcircled{b} does not show any horizontal nodal lines. Since the first two lowest energy branches share the same upward curvature, only indirect bandgap originating in the same BZ is possible. The third lowest energy branch of a BZ which originates in its two neighbouring BZs (aided by zone folding) has downward curvature. Thus, only a direct bandgap can be supported between this and the second lowest energy curve at the BZ boundary.

\begin{figure}[!ht]
\includegraphics[width=8.5 cm]{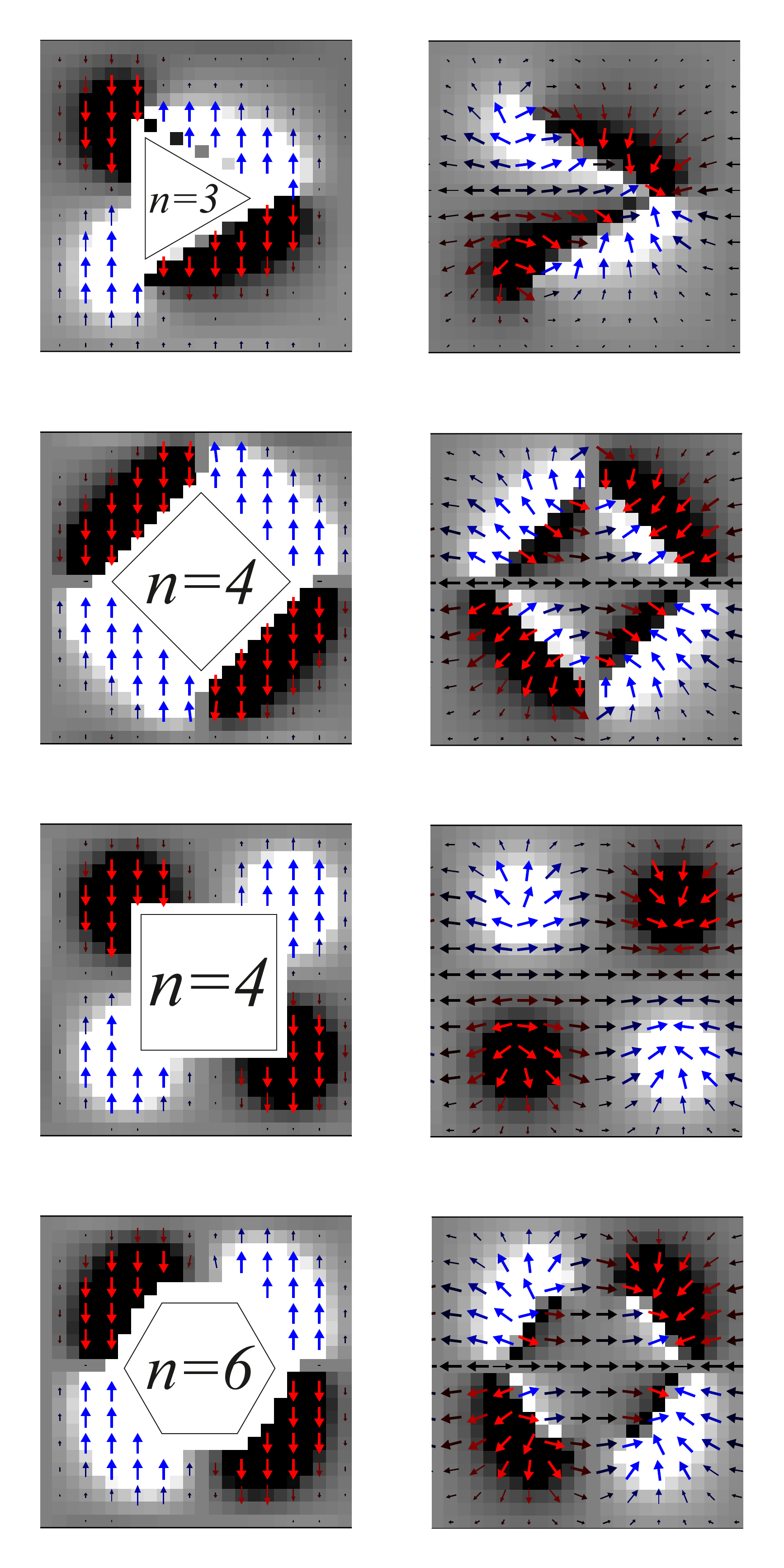}
\caption{(Colour Online) Exchange (left column) and demagnetization (right column) field profiles at $t=0$ for $n=3, 4$ \& $6$ (marked by insets).}
\label{Fig:field}
\end{figure}

A quick visual comparison of different dispersion relations displayed in Fig. \ref{Fig:dispersion} reveals a qualitative convergence of dispersion modes starting as early as $n=4$ (square antidots). No new band gaps open or close. Reference \onlinecite{Klos2012} talks about such similarities between results from square and circular antidot based MAWs and how this convergence, or insensitivity towards the shape of the hole is desirable for the functioning of MAWs. However, note that when the square antidots are tilted by $45^{\circ}$ (diamond shaped antidots) (see Fig. \ref{Fig:dispersion}, left column middle row), one of the band gaps from $n=3$ case is partially restored. The computations of the exchange and the dipole field profiles (EFPs and DFPs) are done to help understand the cause for this observation. These profiles are shown in Fig. \ref{Fig:field}. It may be noted how the EFP around the square antidots matches to that around the hexagonal antidots. They have similar field orientations and cover similar regions in space. Maximum value of the this field is of the order of $20\%$ of $M_s$. However, their demagnetizing field profiles do not match well. On the other hand, the demagnetizing field profile around the tilted square antidots matches better with the same around the hexagonal antidots (similar field orientations and elongated coverage in space and comparable maxima of the order of $50\%$ of $M_s$). Hence, the demagnetizing field or its corresponding potential distribution, may not be the cause of the observed changes in the band structure. Dipole dominated SWs, which occur in much larger structural dimensions are more likely to be affected by the demagnetizing field distribution. To further test the postulate, that the dispersion in considered MAWs is largely dependent upon the exchange field distribution, the case of diamond shaped antidots was considered. It was anticipated that these antidots will produce elongated regions of inhomogeneous exchange fields (similar to what is observed along the slanting edges of the triangular antidots) as opposed to chiefly circular ones (which is seen in the case of square antidots). Surly enough, the exchange field profiles of triangular, diamond shaped and square antidots were remarkably different from each other (as one of the edges of triangular antidot is vertical). This establishes a correlation of observed SW dispersion on their exchange instead of their demagnetizing field distribution.

Exchange energy density, $E_{\text{exch}}\left(\bf{r}_i\right)$, which contributes to the total energy $\bf{M}\cdot\bf{H}_{\text{eff}}$, is isotropic in a homogeneous magnetic medium with uniform exchange coefficient $A$. This field is calculated in OOMMF \cite{Donahue2002} as given below:

\begin{eqnarray}
E_{\text{exch}}\left({\bf r}_i\right)=A{\bf m}\left({\bf r}_i\right)\cdot\sum_{{\bf r}_j}\frac{{\bf m}\left({\bf r}_i\right)-{\bf m}\left({\bf r}_j\right)}{\left|{\bf r}_i-{\bf r}_j\right|^2}\text{.}
\label{eq:exchange}
\end{eqnarray}

Where ${\bf r}_j$ enumerates the region in the immediate neighbourhood of ${\bf r}_i$. In the absence of SW dynamics ${\bf m}({\bf r}_i)-{\bf m}({\bf r}_j){\simeq}0$ except where ${\bf r}_j$ lies close to antidot boundary. Therefore, by changing its geometrical boundary, the exchange field distribution around an antidot can be changed. This can conceivably scatter exchange dominated SWs differently and alter their resultant dispersion relation.

It also needs to be considered if the simulations represent the physical reality. Particularly, how can FDM or FEM based ordinary differential equation solvers like OOMMF or Nmag, which necessarily discretize the continuous sample, calculate the isotropic exchange energy and the demagnetization energy \cite{Donahue2007} with good accuracy? Reference \onlinecite{Donahue1997} concludes that the discrete representations should yield accurate results for ${\pi}d/a={\pi}/24\text{ }{\ll}\text{ }1$. This was further confirmed by the fact that using $d=0.5$ nm for the MAW with tilted square antidots did not alter the exchange field distribution significantly.

\section{Conclusions}\label{Sec:IV}

We have discussed the dispersion of spin-waves in nanoscale one-dimensional magnonic antidot waveguides. In particular we have observed how an antidot's geometry can affect the said dispersion. By dint of power and phase distribution profiles of different spin-wave modes, we have explored the origin of direct and indirect bandgaps that were encountered in the obtained dispersion relations. This understanding can be used, for example, to more readily design for the direct bandgaps and avoid the indirect ones. We have also studied the degree and nature of the inhomogeneity in the exchange field distribution around the edges of an antidot. Apart from offering a way to control the band structure of the exchange dominated spin-waves, we have also demonstrated their dependence on the exchange field profile around the antidots. We demonstrated that useful direct bandgaps can be opened at the same filling fraction without removing additional material during fabrication. Demagnetizing field profile, whose intensity here reached over $0.5 M_{\text{s}}$, is expected to affect the dispersion relations on (thousand times) greater length scales. Without considering the changes in the exchange field distribution, the same has been established by Ref. \onlinecite{Klos2012-2} in two-dimensional magnonic crystals where the hole is filled up by another magnetic material. However, forbiddingly vast computational resources will be required to obtain those results with good frequency and wavevector domain resolutions without compromising the accuracy of the dynamics.

\begin{acknowledgements}
We acknowledge the financial support from the Department of Science and Technology, Government of India (grant no. INT/EC/CMS 24/233552), Department of Information Technology, Government of India (grant no. 1(7)/2010/M\&C), the European Community's Seventh Framework Programme FP7/2007-2013 (Grant Agreement no. 233552 for DYNAMAG and Grant Agreement no. 228673 for MAGNONICS) and the EPSRC Doctoral Training Centre Grant EP/G03690X/1. D. K. would like to acknowledge financial support from CSIR - Senior Research Fellowship (File ID: 09/575/(0090)/2011 EMR-I) and useful discussions with M. Donahue,  M. Krawczyk and J. W. K{\l}os. 
\end{acknowledgements}


\end{document}